# Maximally-localized Wannier functions for GW quasiparticles

D. R. Hamann[1,2] and David Vanderbilt[1]

[1]Department of Physics and Astronomy, Rutgers University, Piscataway, NJ 08854-8019
[2]Mat-Sim Research LLC, P. O. Box 742, Murray Hill, NJ, 07974

## ABSTRACT

We review the formalisms of the self-consistent GW approximation to many-body perturbation theory and of the generation of optimally-localized Wannier functions from groups of energy bands. We show that the quasiparticle Bloch wave functions from such GW calculations can be used within this Wannier framework. These Wannier functions can be used to interpolate the many-body band structure from the coarse mesh of Brillouin zone points on which it is known from the initial calculation to the usual symmetry lines, and we demonstrate that this procedure is accurate and efficient for the self-consistent GW band structure. The resemblance of these Wannier functions to the bond orbitals discussed in the chemical community led us to expect differences between density-functional and many-body functions that could be qualitatively interpreted. However, the differences proved to be minimal in the cases studied. Detailed results are presented for $SrTiO_3$ and solid argon.

# I. INTRODUCTION

For several decades, many-body perturbation theory has been used successfully to describe excited-state electronic properties of a broad range of solids. The spectra of excited electrons and holes in solids are properties of the single-particle Green's function, which is determined by the electron kinetic energy, the ionic and Hartree potentials, and the self-energy operator $\Sigma$ which encompasses all the electron-electron exchange and correlation effects. While $\Sigma$ cannot be calculated exactly, an approximation $\Sigma \approx GW$, where $G$ is the Green's function and $W$ is the dynamically screened Coulomb interaction, was proposed by Hedin in 1965.[1] Two decades of development of electronic-structure calculations within the local-density approximation (LDA)[2] would pass, however, before full-blown *ab-initio* implementations of this so-called GW approximation would be realized.[3,4]

Widespread application of GW calculations to metals, semiconductors, and insulators gave good agreement with experimental band structures in many cases. In general, these calculations followed the pioneering works in using LDA eigenvalues and Bloch functions to evaluate G and W, and to find the so-called quasiparticle eigenvalues from the diagonal expectation value of the $\Sigma$ operator and its energy derivative, also calculated with LDA Bloch functions.[3,4] In principle, it is desirable to evaluate the GW approximation self-consistently, since different results would be expected if a different mean-field approximation such as Hartree-Fock were used instead of LDA in this "one-shot" scheme. $\Sigma$ is a non-Hermitian and energy-dependent operator, reflecting the fact that the spectral weight of G contains a broad continuum representing many-particle excitations as well as relatively sharp quasiparticle peaks. This complicates the issue of self-consistency within a set of independent-particle-like quasiparticle wave functions. A recently introduced approximation to quasiparticle self-consistent GW (QSGW) generates such wave functions as eigenfunctions of a Hamiltonian containing a Hermitian, time-independent effective exchange-correlation potential $V_{xc}^{eff}$ constructed from $\Sigma$.[5,6] This potential was subsequently shown to approximately minimize a plausibly-defined measure of the difference of the time-evolution determined by it and by the full $\Sigma$.[7] QSGW band gaps showed significant improvement over those obtained from one-shot LDA-GW band gaps for a variety of materials.[7]

One practical problem presented by QSGW calculations is that $V_{xc}^{eff}$ is not simply a potential but a non-local operator. After the calculations have been iterated to self-consistency, $V_{xc}^{eff}$ is defined only on the uniform mesh of Brillouin-zone **k** points, so there is no straightforward way to calculate quasiparticle eigenvalues at arbitrary **k** points such as those along the symmetry lines used to plot band structures. This was one of two issues motivating the present study.

The authors introducing the QSGW method have further argued that the QP wave functions obtained by their method are physically meaningful representations of the correlated quasiparticle states envisaged in the Landau quasiparticle picture, and can be used in the calculation of physical properties.[6] The density calculated from these wave functions, for example, is used to calculate the Hartree potential in their QSGW procedure. It was recently shown that such QP Bloch functions can differ substantially from their LDA counterparts, especially at general points in the Brillouin zone.[8] The shape of individual Bloch functions is not easily assigned a physical interpretation, however, and it is this issue which provides the second motivation for the present study.

An alternative to representing one-particle-like electronic states of solids as periodic Bloch functions $\psi_{\mathbf{k}n}(\mathbf{r})$ is to represent them as localized Wannier functions $w_n(\mathbf{r}-\mathbf{R})$, where $n$ is a band index and $\mathbf{R}$ is the lattice vector of a unit cell.[9] The original concept of Wannier functions associated with single isolated bands has been generalized to sets of Wannier functions associated with isolated groups of bands, in the process introducing an algorithm to minimize their spatial spread.[10] A further generalization permitted these maximally localized Wannier functions (MLWF's) to be constructed from entangled bands.[11] As we will discuss in more detail below, the construction of MLWF's requires a set of Bloch functions on a uniform mesh of $\mathbf{k}$ points, which is precisely what we have for the QP functions at the end of the QSGW calculation. MLWF's form a basis set that can be used to generate a highly accurate interpolated band structure at very low computational cost,[11] thereby solving the basic practical difficulty of QSGW calculations discussed above.[12]

MLWF's turn out to be the solid-state equivalent of the localized molecular orbitals studied in the chemical literature.[13] These linear combinations of delocalized molecular-orbital eigenfunctions correspond closely to the "natural bond orbitals" which form a realization of the chemists' picture of localized bonds and lone pairs as basic units of molecular structure.[14] Comparing LDA and QP MLWF's could potentially offer qualitative insight into the manner in which an improved treatment of many-body correlations alters individual bonds. In effect, MLWF's could extract the physical content of changes observed in individual Bloch functions.[8]

A third potentially interesting comparison relates to the modern theory of electric polarization in solids.[15,16,17] The theory was originally formulated in terms of a Berry phase,[18] but it can be shown to be equivalent of calculating the centers of charge of Wannier functions.[16] While the Berry phase formulation has been formally extended to include many-body wave functions,[19] the connection to QP Wannier centers through the GW approximation and the single-particle-like QP Bloch functions considered here[6] has not been explored, and is beyond the scope of the present investigation.

## II. FORMALISM AND IMPLEMENTATION

### A. Quasiparticle self-consistent GW

In this section we shall briefly review the mathematical expressions defining QSGW in the approximation of Refs. 5-7 without discussion of the underlying rationale for this or the overall GW approximation, for which we refer the readers to these and Refs. 1, 3 and 4. The single-particle-like QP wave functions we have been discussing are solutions of the effective Schrödinger equation

$$H^{eff}\psi_{\mathbf{k}i}(\mathbf{r}) = \int d^3r' \left\{ \left[ -\tfrac{1}{2}\nabla^2 + V_{Hartree}(\mathbf{r}) \right] \delta(\mathbf{r}-\mathbf{r}') \right.$$
$$\left. + V_{Ext}(\mathbf{r},\mathbf{r}') + V_{xc}^{eff}(\mathbf{r},\mathbf{r}') \right\} \psi_{\mathbf{k}i}(\mathbf{r}') = \varepsilon_{\mathbf{k}i}\psi_{\mathbf{k}i}(\mathbf{r}) \quad (1)$$

where we have written the external potential as a non-local operator with pseudopotentials in mind, and where the Hartree and effective exchange-correlation potentials are to be determined self-consistently. The Hartree potential is calculated from the density defined in the usual way from the occupied $\psi_{\mathbf{k}i}$.

What we will call the effective Green's function is computed from the solutions of this equation as

$$G_{eff}(\mathbf{r},\mathbf{r}',\omega) = \frac{V}{(2\pi)^3} \int_{BZ} d^3k \sum_{\text{Bands } i} \frac{\psi_{\mathbf{k}i}^*(\mathbf{r})\psi_{\mathbf{k}i}(\mathbf{r}')}{\omega - \varepsilon_{\mathbf{k}i} + i\delta\,\text{sgn}(\varepsilon_{\mathbf{k}i})} \quad (2)$$

where $V$ is the unit cell volume, $\delta$ is a positive infinitesimal and the QP energies $\varepsilon_{\mathbf{k}i}$ are measured from the Fermi energy.

The dynamically screened interaction $W$ is calculated within the random phase approximation (RPA) starting with the susceptibility

$$\chi(\mathbf{r},\mathbf{r}',\omega) = \frac{V}{(2\pi)^3} \int_{BZ} d^3k \sum_{ij} (n_{\mathbf{k}i} - n_{\mathbf{k}j}) \frac{\psi_{\mathbf{k}i}(\mathbf{r})\psi_{\mathbf{k}j}^*(\mathbf{r})\psi_{\mathbf{k}i}^*(\mathbf{r}')\psi_{\mathbf{k}j}(\mathbf{r}')}{\varepsilon_{\mathbf{k}i} - \varepsilon_{\mathbf{k}j} - \omega - i\delta} \quad (3)$$

where $n_{\mathbf{k}i}$ and $n_{\mathbf{k}j}$ are occupation numbers. The RPA dielectric function is then given as

$$\varepsilon(\mathbf{r},\mathbf{r}',\omega) = \delta(\mathbf{r}-\mathbf{r}') - \int d^3r'' V_c(\mathbf{r},\mathbf{r}'')\chi(\mathbf{r}'',\mathbf{r}',\omega) \quad (4)$$

where $V_c$ is the Coulomb interaction $1/|\mathbf{r}-\mathbf{r}''|$. Finally the inverse dielectric function $\varepsilon^{-1}$ found as the solution of the integral equation

$$\int d^3r'' \varepsilon(\mathbf{r},\mathbf{r}'',\omega)\varepsilon^{-1}(\mathbf{r}'',\mathbf{r}',\omega) = \delta(\mathbf{r}-\mathbf{r}') \quad (5)$$

yields $W(\mathbf{r},\mathbf{r}',\omega) = \varepsilon^{-1}V_c$.

The self-energy operator $\Sigma$ is given in terms of the above functions as the convolution

$$\Sigma(\mathbf{r},\mathbf{r}',\omega) = \frac{i}{4\pi} \int_{-\infty}^{\infty} d\omega' W(\mathbf{r},\mathbf{r}',\omega')G_{eff}(\mathbf{r},\mathbf{r}',\omega+\omega'). \quad (6)$$

The "full" Green's function $G$ whose spectral weight would consist of both a quasiparticle pole with weight less than one and an incoherent continuum could be calculated from this $\Sigma$ by solving the Dyson equation written schematically as $G = G_{eff} + G_{eff}\Sigma G$, but this is not required in the present QSGW scheme. Instead, a measure of the "distance" between the time evolution generated by $\Sigma$ and that generated by $V_{xc}^{eff}$ is introduced,[5,7]

$$M\left[V_{xc}^{eff}\right] = \int_{BZ} d^3k \sum_i \int_{-\infty}^{\infty} d\omega \langle \psi_{\mathbf{k}i} | \left[ (\Sigma - V_{xc}^{eff})\delta(\omega - H^{eff})(\Sigma - V_{xc}^{eff})^\dagger + h.c. \right] | \psi_{\mathbf{k}i} \rangle. \quad (7)$$

This positive-definite distance measure is approximately minimized by setting[5,7]

$$V_{xc}^{eff}(\mathbf{r},\mathbf{r}') = \frac{V}{4(2\pi)^3} \int_{BZ} d^3k \sum_{ij} \psi_{\mathbf{k}i}^*(\mathbf{r}) \left[ \Sigma_{\mathbf{k}ij}(\varepsilon_{\mathbf{k}i}) + \Sigma_{\mathbf{k}ji}^*(\varepsilon_{\mathbf{k}i}) + \Sigma_{\mathbf{k}ij}(\varepsilon_{\mathbf{k}j}) + \Sigma_{\mathbf{k}ji}^*(\varepsilon_{\mathbf{k}j}) \right] \psi_{\mathbf{k}j}(\mathbf{r}'), \quad (8)$$

where

$$\Sigma_{\mathbf{k}ij}(\omega) = \langle \psi_{\mathbf{k}i} | \Sigma(\mathbf{r},\mathbf{r}',\omega) | \psi_{\mathbf{k}j} \rangle. \quad (9)$$

Eq. (8) completes the self-consistency loop. In practice, Eqs. (1)-(9) are solved iteratively starting with an approximation such as LDA for $V_{xc}^{eff}$. While we have outlined the formalism in real space, most of the indicated operations are in practice carried out in a reciprocal-space representation, and the indicated **k** integrals carried out as sums on a uniform grid.[3,4]

## B. Wannier functions for quasiparticles

A set of $N$ generalized Wannier functions $w_{\mathbf{R}i}(\mathbf{r})$ labeled by index $i$ and lattice vector **R** are constructed as

$$w_{\mathbf{R}i}(\mathbf{r}) = \frac{V}{(2\pi)^3} \int_{BZ} d^3k \, e^{-i\mathbf{k}\cdot\mathbf{R}} \sum_{n=1}^{N_\mathbf{k}} U_{ni}^{(\mathbf{k})} \psi_{\mathbf{k}n}(\mathbf{r}) \quad (10)$$

from Bloch functions $\psi_{\mathbf{k}n}$ with energies inside an energy window including $N_\mathbf{k} \geq N$ bands throughout the BZ, where $V$ is the volume of the unit cell and the $N_\mathbf{k} \times N$ matrices $U_{ni}^{(\mathbf{k})}$ are to be determined. For Wannier functions constructed from an isolated set of bands, $N_\mathbf{k} = N$ for all **k**, and $U_{ni}^{(\mathbf{k})}$ are required to be unitary, but this still leaves a great deal of freedom in their choice. A physically reasonable choice is to require these generalized Wannier functions to be as local as possible. A measure of their locality which is the exact analogue of a criteria of Boys[13] for the molecular-orbital case is the sum $\Omega$ of second moments of the corresponding Wannier functions,

$$\Omega = \sum_{i=1}^{N} \left[ \langle w_{\mathbf{0}i} | r^2 | w_{\mathbf{0}i} \rangle - | \langle w_{\mathbf{0}i} | \mathbf{r} | w_{\mathbf{0}i} \rangle |^2 \right], \quad (11)$$

where we can specialize to the unit cell at the origin, since all sets of Wannier functions are equivalent within a lattice-vector translation, $w_{\mathbf{R}i}(\mathbf{r}) = w_{\mathbf{0}i}(\mathbf{r}-\mathbf{R})$.

The when Eq. (10) is substituted in Eq.(11), $\Omega$ becomes a function of the $U$ matrices and matrix elements of **r** and $r^2$ between pairs of Bloch functions $\psi_{\mathbf{k}n}$. These matrix elements can be reexpressed as matrix elements of gradients and Laplacians with respect to the Bloch **k** vector. Since the **k** integrations in Eq. (10) are, as usual, approximated by Brillouin zone sums on uniform grids, finite-difference expressions for these gradients and Laplacians can be formulated in terms of matrix elements

$$M_{mn}^{(\mathbf{k},\mathbf{b})} = \langle \psi_{\mathbf{k}m} | e^{-i\mathbf{b}\cdot\mathbf{r}} | \psi_{\mathbf{k}+\mathbf{b}n} \rangle \quad (12)$$

where the set of vectors $\{\mathbf{b}\}$ connect each **k**-space mesh point with its nearest neighbors. With $\Omega$ expressed in terms of $M_{mn}^{(\mathbf{k},\mathbf{b})}$ and $U_{ni}^{(\mathbf{k})}$, and specializing to the unitary case $N_{\mathbf{k}} = N$, it is possible to calculate the derivatives of $\Omega$ with respect to the $U_{ni}^{(\mathbf{k})}$, and use this as the basis of an algorithm to minimize $\Omega$ as discussed in detail in Ref. 10.

It is desirable to start the minimization algorithm from an initial approximation to $U_{ni}^{(\mathbf{k})}$ which is based on some physically-motivated picture of orbitals or bonds that one expects to be associated with the set of bands being considered. $N$ guiding functions $g_i(\mathbf{r})$ having appropriate centers and orbital characters (e.g., single-gaussian atomic-like orbitals with s, p, or d angular dependence, or hybrid combinations such as sp$^3$) are introduced. The overlaps with the Bloch functions are computed,

$$A_{ni}^{(\mathbf{k})} = \langle \psi_{\mathbf{k}n} | g_i \rangle, \tag{13}$$

and the $A_{ni}^{(\mathbf{k})}$ matrices used in combination with a symmetric orthonormalization procedure to form a starting approximation to $U_{ni}^{(\mathbf{k})}$. Since the $U$'s at each $\mathbf{k}$ are coupled to those at neighboring $\mathbf{k}$'s through the $M_{mn}^{(\mathbf{k},\mathbf{b})}$, the minimization must be solved self-consistently throughout the Brillouin zone and this algorithm proceeds iteratively, updating the set of $U$'s at each step. When no further significant reduction in $\Omega$ can be obtained, maximally localized Wannier functions will have been constructed.[10]

When the bands possessing the orbital character of interest do not occur as an isolated group, the bands are said to be entangled and the energy window must be chosen so that $N_{\mathbf{k}} \geq N$ for at least some $\mathbf{k}$. For the isolated-group case, $\Omega$ can be divided into two positive-definite terms, $\Omega = \Omega_I + \tilde{\Omega}$. The so-called invariant term $\Omega_I$ can be calculated directly from the $M_{mn}^{(\mathbf{k},\mathbf{b})}$ and is not dependent on the $U$'s or changed by the optimization algorithm. This forms the basis for the extension of the MLWF procedure to the entangled case. For each $\mathbf{k}$, we generate $N$ orthonormal $\phi_{\mathbf{k}j}$ as linear combinations of the $N_{\mathbf{k}}$ $\psi_{\mathbf{k}n}$ within the window,

$$\phi_{\mathbf{k}m}(\mathbf{r}) = \sum_{n=1}^{N_{\mathbf{k}}} D_{nm}^{(\mathbf{k})} \psi_{\mathbf{k}n}(\mathbf{r}), m = 1, N, \tag{14}$$

where $D^\dagger D = 1$. For those $\mathbf{k}$ at which $N_{\mathbf{k}} = N$, the "disentanglement" matrix $D$ is simply the unit matrix. $\Omega_I$ is calculated as for the isolated-group case using the overlap matrix for neighboring $\phi$'s, $M'^{(\mathbf{k},\mathbf{b})} = \left(D^{(\mathbf{k})}\right)^\dagger M^{(\mathbf{k},\mathbf{b})} D^{(\mathbf{k}+\mathbf{b})}$. A new optimization algorithm is introduced to minimize $\Omega_I$ with respect to the $D_{nm}^{(\mathbf{k})}$. Since the $D$'s at each $\mathbf{k}$ are also coupled to those at neighboring $\mathbf{k}$'s through the $M_{nm}^{(\mathbf{k},\mathbf{b})}$, this minimization must also be solved self-consistently throughout the Brillouin zone by iteration.[11]

After this minimization has converged, it is desirable to diagonalize the set of $N \times N$ Hamiltonians $H'$ in the $\phi$ subspaces

$$H_{mn}'^{(\mathbf{k})} = \sum_{\ell=1}^{N_{\mathbf{k}}} D_{\ell m}^{*(\mathbf{k})} \varepsilon_{\mathbf{k}\ell} D_{\ell n}^{(\mathbf{k})}, \tag{15}$$

yielding eigenvalues $\tilde{\varepsilon}_{\mathbf{k}n}$, and eigenvectors from which Bloch functions $\tilde{\psi}_{\mathbf{k}n}$, which are linear combinations of the $\phi_{\mathbf{k}m}$, and transformed matrices $\tilde{D}^{(\mathbf{k})}_{nm}$ can be constructed. The Wannier functions are now expressed as

$$w_{\mathbf{R}i}(\mathbf{r}) = \frac{V}{(2\pi)^3} \int_{BZ} d^3k\, e^{-i\mathbf{k}\cdot\mathbf{R}} \sum_{n=1}^{N} \tilde{U}^{(\mathbf{k})}_{ni} \tilde{\psi}_{\mathbf{k}n}(\mathbf{r}) \qquad (16)$$

and $\tilde{\Omega}$ is calculated from the $\tilde{\psi}_{\mathbf{k}n}$ overlaps, $\tilde{M}^{(\mathbf{k},\mathbf{b})} = \left(\tilde{D}^{(\mathbf{k})}\right)^{\dagger} M^{(\mathbf{k},\mathbf{b})} \tilde{D}^{(\mathbf{k}+\mathbf{b})}$, and minimized with respect to the matrix elements of the set of unitary $\tilde{U}$'s using the isolated-group algorithm. When this minimization has converged, the original $U$'s in Eq.(10) are simply the set of matrix products $U = \tilde{D}\tilde{U}$.

It is often desirable to limit the mixing of Bloch functions in Eq.(14) so that the $\psi_{\mathbf{k}n}$ belonging to some bands (eg., low-lying conduction bands) can be exactly reproduced by linear combinations of $w_{\mathbf{R}i}$. In this case, we introduce another energy window within the overall outer window which we will call the frozen window, and constrain the $D$'s so that $D^{(\mathbf{k})}_{nm} = \delta_{nm}$ for $\varepsilon_{\mathbf{k}m}$ within this window.

Band interpolation based on Wannier functions is a form of Slater-Koster tight-binding interpolation,[20] but with Hamiltonian matrix elements calculated directly from the eigenvalues and eigenfunctions of the underlying *ab initio* calculation rather than through a fitting procedure. The first step is to rotate the diagonal Hamiltonian $\tilde{H}^{(\mathbf{k})}_{mn} = \tilde{\varepsilon}_m \delta_{mn}$ at each $\mathbf{k}$ mesh point into the linear combination needed to form the MLWF's,

$$\tilde{H}^{(\mathbf{k},\text{rot})}_{ij} = \sum_{n=1}^{N} \tilde{U}^{*(\mathbf{k})}_{ni} \tilde{U}^{(\mathbf{k})}_{nj} \tilde{\varepsilon}_{\mathbf{k}n}. \qquad (17)$$

We have used the quantities with tildes introduced for the disentangling case without loss of generality, since they reduce to their "untilded" counterparts for an isolated group of bands. The next step is to form the matrix elements of the Hamiltonian between origin-based MLWF's and those at a set of lattice vectors $\mathbf{R}$ within a Wigner-Seitz supercell centered at the origin and chosen so that the number of $\mathbf{R}$'s equals the number of points in the $\mathbf{k}$ mesh, $N_{\text{kp}}$. This is carried out by the discrete Fourier transform

$$\tilde{H}^{(\mathbf{R},0)}_{ij} = \langle w_{\mathbf{R}i} | \tilde{H} | w_{\mathbf{0}j} \rangle = \frac{1}{N_{\text{kp}}} \sum_{\mathbf{k}\,\text{mesh}} e^{-i\mathbf{k}\cdot\mathbf{R}} \tilde{H}^{(\mathbf{k},\text{rot})}_{ij}. \qquad (18)$$

Finally, for any arbitrary point $\mathbf{k}'$ we find and diagonalize the $N \times N$ Hamiltonian

$$\tilde{H}^{(\mathbf{k}',\text{rot})}_{ij} = \sum_{\mathbf{R}} e^{i\mathbf{k}'\cdot\mathbf{R}} \tilde{H}^{(\mathbf{R},0)}_{ij}, \qquad (19)$$

thereby obtaining the interpolated energies $\tilde{\varepsilon}_{\mathbf{k}'n}$ by an exceedingly fast computation. We note that the energies obtained in this manner must be identical to the input $\tilde{\varepsilon}_{\mathbf{k}n}$ when $\mathbf{k}'$ lies on a mesh point $\mathbf{k}$, and equal to $\varepsilon_{\mathbf{k}n}$ within a frozen window or for an isolated set of bands. Very accurate interpolated band structures have been demonstrated by this method for the LDA case, where the complete *ab initio* band structure is easily calculated for comparison.[10,11]

We note that good interpolation requires Wannier functions whose individual spreads, given by the square roots of each term in the sum in Eq.(11), are small compared to the size of the supercell. $\tilde{H}_{ij}^{(\mathbf{R},\mathbf{0})}$ as calculated by Eq.(18) is actually the sum of matrix elements $\langle w_{(\mathbf{R}+\mathbf{R}_{SC})i} | \tilde{H} | w_{\mathbf{0}j} \rangle$ over the Bravais vectors $\mathbf{R}_{SC}$ defining the superlattice. If contributions from $\mathbf{R}_{SC} \neq \mathbf{0}$ are significant, spurious oscillations of the bands along lines in **k** space could be generated by the inverse discrete Fourier transform in Eq.(19). In some cases, this problem may only be solved by choosing a denser **k** mesh for the calculation, and hence a larger supercell.

Throughout this section, we have made no distinction between LDA and QSGW Bloch functions and energies. There is none, since the MLWF construction algorithms are driven solely by overlaps, guiding functions, eigenvalues, and choice of energy windows without regard to the physical approximations or mathematical forms leading to the Bloch functions (such as local vs. non-local $V_{xc}$ operators).

**C. Implementation**

The computation of MLWF's for QSGW quasiparticles had been implemented utilizing two existing publicly-available computer codes. The ABINIT package is a full-featured implementation of density functional theory and density functional perturbation theory based primarily on pseudopotentials and a plane-wave basis set.[21] It has been extended to include self-consistent GW capabilities[8,22] within the QSGW framework.[5-7] The Wannier-function algorithms described in Refs. 10 and 11 are implemented in the WANNIER90 package.[23] This package includes both a stand-alone program which needs a set of files produced by an *ab-initio* program, and a library whose routines can be called from within another program.

At the time the present project was undertaken, an interface of ABINIT and the WANNIER90 library was partially completed for density-functional wave functions. We have substantially extended the capabilities of this interface, a principal addition being the implementation of a set of guiding functions and generation of the corresponding *A* matrices of Eq.(13) with the full set of features specified in Tables 3.1 and 3.2 of the WANNIER90 User Guide,[23,24] allowing atomic-like orbitals and hybrids to be centered at arbitrary sites and oriented along arbitrary axes.

For reasons of efficiency, the implementation of QSGW in ABINIT uses LDA Bloch functions as a basis set for the expansion of the quasiparticle Bloch functions rather than calculating these directly in the underlying plane-wave basis.[8] To avoid duplication and dealing with detailed differences in wave function storage in the density-functional and GW sections of ABINIT, the density-functional-Wannier interface was retained for the quasiparticle MLWF calculations. The quasiparticle eigenvectors in the LDA basis form a unitary transformation which is updated and saved after each iteration of a QSGW calculation. To adapt the interface to GW quasiparticles, it is merely necessary to apply this unitary matrix to the LDA Bloch basis functions prior to generating the *M* and *A* matrices of Eqs.(12) and (13), and to generating data for plotting the Wannier functions.[24]

### III. RESULTS AND DISCUSSION

LDA and QSGW calculations were carried out for several systems to test the methods reported here and to explore the differences of the MLWF's in the two approximations. A one-sentence summary that will satisfy the disinterested reader is that the differences are extremely small and that band interpolation works as well for GW as for LDA bands.

All calculations were carried out using norm-conserving pseudopotentials generated from LDA atomic calculations.[25] The GW calculations utilized dielectric matrices, Eq.(4), represented in the generalized-plasma-pole approximation.[3] There are many choices of Brillouin-zone meshes, wave vector cutoffs, and numbers of bands to be utilized in various portions of the GW calculations, and we explored the convergence of our results in sufficient detail to believe that our GW band energies were converged to within ~0.1eV. Four iterations of the self-consistency loop generally sufficed at this level.

The first system explored was Si in the usual diamond structure. MLWF's were generated for both LDA and GW results with guiding functions to select either 4 bonding MLWF's for the valence bands and 4 antibonding ones for the low-lying conduction bands, or 8 sp3-like MLWF's for all these bands. Isosurface plots of these functions were very similar to those of Fig. 9 of Ref.11, and LDA and GW MLWF's were virtually indistinguishable for any reasonable choice of isosurface amplitude. Band interpolation results for LDA were similar in their ability to reproduce full *ab initio* bands to the comparison shown in Fig. 8 of Ref.11. The GW valence bands were very similar to those of the LDA, and the GW conduction bands closely approximated a rigid upward shift of the LDA bands. Our minimum gap increased from 0.49eV to 1.36eV, compared to 1.15eV (experiment[26]), 1.47eV (plane-wave pseudopotential QSGW[8,27]), and 1.25eV (linear muffin-tin orbital QSGW[6,27]).

Speculating that the increased GW gap might shift the character of a polar semiconductor's wave functions to appear less covalent and more ionic, we next explored AlP, which has the zincblende structure. Once again, isosurface plots were essentially indistinguishable. Counterintuitively, the center of the bonding MLWF's shifted 0.007Å towards the Al, and an isosurface plot constructed to greatly exaggerate the GW-LDA difference showed a very slightly more covalent character for GW. Band interpolation results were as described for Si, with the minimum gap raised from 1.49eV to 2.76eV, compared to 2.51eV(experiment[28]) and 2.61eV (LMTO QSGW[29]).

Another system we explored was a cubic perovskite version of $SrZrS_3$ (whose real structure is a distorted perovskite[30]). This was chosen as a computationally less demanding analogue of SrTiO$_3$, whose highly localized orbitals require high plane-wave cutoff energies. Once again, our expectation was reduced covalency between the S 3p and Zr 4d orbitals, and once again our MLWF results showed extremely small differences. While we expected to generate MLWF's for the low-lying conduction bands based on just the Zr 4d's, these bands were sufficiently entangled that the algorithms failed to converge to satisfactory results, and we didn't pursue this further by introducing more bands and more Wannier functions.

In attempting to find more significant differences in LDA and GW MLWF's, we turned to SrTiO$_3$ itself, and finally to solid Ar, whose conduction-band Bloch functions were reported to display large differences in Ref. 8. These results will be discussed in detail in the following subsections.

**A. SrTiO$_3$**

We studied cubic SrTiO$_3$ using the experimental lattice constant of 3.905Å.[31] The plane-wave energy cutoffs employed were 60Ry for the wave functions and 25Ry for the dielectric function. A basis of LDA Bloch functions for 80 bands on a $\Gamma$−centered $8\times8\times8$ **k**-mesh was used for the QSGW calculation, with 27 bands treated self-consistently.[32] 25 unoccupied bands were used in the dielectric function calculation. In constructing norm-conserving pseudopotentials for this calculation, the only semi-core states treated as valence were the Sr 4p, which are nearly degenerate with the O 2s.[33]

The MLWF's were constructed in two groups. The first group was generated from s and p guiding functions on the three O's, and p's on the Sr. The energy window included all the valence bands, which formed an isolated group. Isosurface plots of the predominantly O p$_z$ and p$_x$ GW MLWF's are shown in Fig. 1. They show some covalent sigma (p$_z$) and pi (p$_x$) bonding with Ti e$_g$ and t$_{2g}$ d orbitals, respectively, as seen in earlier LDA MLWF's for similar perovskites.[34] MLWF's for the low-lying conduction bands were generated from Ti d guiding functions, and the LDA and "enhanced" (see below) GW d$_{z^2}$ functions are compared in Fig. 2, which show a small sigma antibonding admixture of O p$_z$.[35]

As was the case for our earlier examples, the SrTiO$_3$ valence and conduction MLWF's show no visually apparent differences between LDA and QSGW for any choice of isosurface value which reasonably displays the shape of the functions. As a quantitative measure of their similarities, we calculated their overlaps by numerical integration on a real-space grid within a $3\times3\times3$ supercell. The overlaps were 0.9995, 0.9997, 0.9983, and 0.9991 for the O p$_x$, O p$_y$, Ti e$_g$, and Ti t$_{2g}$ – like functions, respectively, bearing out the qualitative observations from the plots. Picking the "best case" Ti d$_{z^2}$, we artificially exaggerated the difference by plotting the isosurface for $10w_{GW}(\mathbf{r}) - 9w_{LDA}(\mathbf{r})$ labeled "enhanced" GW in Fig. 2. The antibonding O p$_z$ contribution is seen to be slightly smaller, and while we can speculate that this is related to the increase of the gap, it is obviously a very small effect. The MLWF's of the deep-lying O 2s and Sr 3p bands are compact and atomic-like, and we did not undertake any detailed comparisons of these.

Turning to band interpolation, we display the accuracy with which MLWF interpolation can reproduce the upper valence bands (O 2p) and lower conduction bands (Ti 3d) in Fig. 3. The solid interpolated bands completely obscure the grey dashed lines representing a direct fine-grained LDA calculation on the Brillouin zone symmetry lines except for the uppermost portions of the conduction bands at M and R. The conduction bands are entangled, and the limits of the outer window (OW) and frozen window (FW) are indicated on the band plot. The OW just includes all the Ti 3d-like bands, and the FW was chosen just below the 6$^{th}$ band at $\Gamma$.

Fig. 4 compares the interpolated LDA and GW bands in this same energy range. The dashed GW band lines are in essentially exact agreement with the directly-computed GW energies on the symmetry points which were a part of the **k** mesh, shown as open circles. Choices of energy windows for the GW conduction bands were based on similar criteria to those described above. The gap is essentially doubled from 1.61eV to 3.32eV, is indirect from R to $\Gamma$ in both cases, and can be compared with the experimental gap of 3.16eV.[36] The GW conduction bands are quite well represented by a rigid upward shift of the LDA bands, the total width of the Ti 3d manifold only increasing by 0.37eV. As is clear from Fig. 4, the O 2p valence bands are more significantly broadened, by 0.87eV.

**B. Solid Argon**

The final example we shall report on is solid Ar. Ref. 8 reported significant differences between LDA and QSGW Bloch functions for the Ar conduction bands at general **k** points, as well as a substantial self-consistency correction to the gap. Given the minor differences we have discussed in the other systems, it seemed worthwhile to examine the corresponding Wannier functions, despite the fact that we couldn't expect to say much about bonding, etc. for such states.

Ar has an fcc crystal structure, and we used the experimental lattice constant of 5.31Å.[37] The plane-wave energy cutoffs employed were 40Ry for the wave functions and 32Ry for the dielectric function. A basis of LDA Bloch functions for 30 bands on a $\Gamma-$centered $8\times8\times8$ **k**-mesh was used for the QSGW calculation, with 27 bands treated self-consistently,[32] and 26 unoccupied bands used for the dielectric calculation.

Our first task was to verify that we reproduced the large Bloch function differences. Fig. 5 is modeled on Fig. 9 of Ref. 8, and shows comparably large changes of the second conduction band Bloch function at $\mathbf{k} = (-1/8, -3/8, 1/4)$. The small differences in the absolute values of the our Bloch functions relative to theirs at the "shoulders" near the Ar atoms are likely due to differences in the pseudopotentials. The fact that we used the generalized-plasma-pole approximation[3] for the dielectric function while energy integration[4] was used in Ref. 8 might have some small effect on the GW functions.

The full LDA and Wannier-interpolated LDA conduction band structures are compared in Fig. 6, which follows the conventions of Fig. 3. They are generally consistent with an early augmented-plane-wave calculation.[38] We initially hoped to disentangle and construct MLWF's for the lowest 4 conduction bands, trying a set of 4 sp3 guiding functions centered at the octahedral interstitial site, which seemed a plausible empty site for conduction-band functions whose main characteristic is to be repelled from the Ar by orthogonalization to the core-like valence bands. Experiments with various OW and FW windows typically failed to converge, tended to break the symmetry set by the guiding functions, and occasionally collapsed to quite different functions. Success in obtaining a symmetric, well-localized set of MLWF's was finally achieved by introducing 9 Ar-centered guiding functions, 6 hybrids of s, p, and $e_g$ d functions, and 3 $t_{2g}$ d functions. The frozen-window limit FW was chosen to lie just below the $\Lambda_1$ band emanating from the second $\Gamma_1$ band, since the guiding functions select a Wannier basis which contains only one s function and could not be expected to fit this band. Fits within the frozen window are seen to be excellent, whereas only portions of the band structure in the entangled region above are fit well. In the Wannier fit, bands emanating from $\Gamma_{25'}$ are predominantly composed of the $t_{2g}$ functions, crossing over in most cases to various linear combinations of the hybrids towards other symmetry points.

The QSGW conduction-band MLWF's are shown in Fig. 7. The function (a) was generated from the guiding hybrid $\frac{1}{\sqrt{6}}s + \frac{1}{\sqrt{2}}p_x - \frac{1}{\sqrt{12}}d_{z^2} + \frac{1}{2}d_{x^2-y^2}$, while (b) was generated from $d_{xz}$. The 6 hybrids clearly concentrate much of the weight of the lower bands in an octahedral interstitial site, consistent with our initial speculation. However, the purely $t_{2g}$ d character of the 2$^{nd}$ through 4$^{th}$ bands at $\Gamma_{25'}$ with its dominant weight along nearest-neighbor "bonds" is the probable reason for our lack of success with 4 interstitial-centered functions. We do not show the LDA counterparts of the GW MLWF's because once again, despite the Bloch function differences observed in Ref. 8 and reproduced by our calculations as shown in Fig. 5, the differences are too small to see with any choice of isosurface level which shows the shape of the MLWF's. Quantitatively, the LDA-GW overlaps are 0.9973 for the hybrids and 0.9975 for the $t_{2g}$'s. There seemed little point in showing an "enhanced" function as in Fig. 2 for SrTiO$_3$, since there would be little to say in the way of physical interpretation.

The LDA and QSGW interpolated conduction bands are compared in Fig. 8. As was the case for SrTiO$_3$, the interpolated QSGW bands are in excellent agreement with the directly-calculated ones at the symmetry points contained in the **k** mesh. The energy windows for the MLWF were chosen to parallel the LDA case as closely as possible. Our LDA and QSGW band gaps, 8.13 and 14.49eV respectively, are in good agreement with the values 8.20 and 14.84eV reported in Ref. 8, and with experiment (14.2eV[39]). While it is not immediately apparent from Fig. 8, close inspection shows that the QSGW bands are not rigidly shifted versions of the LDA bands. Shifts within the fitted region range from 6.5 to 9eV, and are not monotonic. We can speculate that this range of shifts could be consistent with the large differences seen between individual LDA and QSGW Bloch functions. However, since the LDA Bloch functions form a very efficient basis for the expansion of the quasiparticle Bloch functions,[8] the individual differences could largely disappear in the sums over the Bloch manifolds forming the Wannier functions.

The valence bands, while wider than previously reported,[38] are quite narrow, and the MLWF's are almost entirely atomic sp-like as expected. We find the 3s band width for LDA and QSGW to be 0.27 and 0.32eV, respectively, and the 3p widths to be 1.33 and 1.53eV. The valence interband splitting increases from 12.98 in LDA to 13.33 in QSGW.

## IV. CONCLUSIONS

We have demonstrated that maximally localized Wannier functions can be formed from quasiparticle wave functions generated using the quasiparticle self-consistent GW approximation to a full many-body treatment of the electronic structure of solids. This was accomplished through relative minor modifications[24] to create the appropriate interface between two publicly-available electronic structure codes, ABINIT[21,22] and WANNIER90.[23]

We have shown through several examples that MLWF interpolation can produce an accurate band structure on the symmetry lines of the Brillouin zone, even though band energies can be directly computed only at few symmetry points through the QSGW calculation itself. While we did not consider any metals as examples, data for accurate Fermi surface plots can be produced by using the WANNIER90 stand-along program to post-process the ABINIT output. Changes in Fermi-surface shape and topology should be expected in comparing LDA and QSGW results.

MLWF's closely correspond to the bond orbitals in terms of which chemists understand bonding in molecules and solids, and we anticipated that changes in the treatment of exchange and correlation would be manifest in the bonding. However, the changes we found in comparing LDA and QSGW functions in fact turned out to be minimal in the examples we studied, even for conduction-band ("antibonding") MLWF's. This was true despite the fact that density-functional theory is formally a ground-state theory, that only densities and not Kohn-Sham wave functions have formal physical significance,[2] and that the QSGW eigenvalues differ significantly from those of the LDA, especially for conduction-band wave functions. Furthermore, large changes had been observed in individual conduction-band Bloch functions,[8] but failed to materialize when many of these were combined to form the Wannier functions. We encourage others to continue this search in hope of finding systems where such changes are large enough to suggest qualitative physical differences found through the many-body approach. Finally, while we have confined out attention to solids, molecules can be treated using the same computational tools by the supercell method, and may prove a more fertile ground for discovering correlation effects not well represented by density functional theory.

**Acknowledgement:** We thank T. Rangel for helpful discussions and for incorporating some of this work in the public release of ABINIT. We thank the XCrysDen project[40] for the isosurface plotting code used to produce some of our figures. DV acknowledges NSF Grant DMR-0549198.

# Figure Captions

Fig. 1 (color online). Isosurface plots of SrTiO$_3$ valence-band maximally localized Wannier functions for GW quasiparticles, at isosurface values $\pm 1/\sqrt{V}$, where $V$ is the unit cell volume, positive values red/light grey, and negative blue/dark grey. (a) is an O-centered p$_z$-like function showing sigma bonding with the Ti d$_{z^2}$ orbital, and (b) an O p$_x$-like function showing pi bonding with Ti d$_{xz}$.

Fig. 2 (color online). Isosurface plots for SrTiO$_3$ conduction-band MLWF's at isosurface values $\pm 2/\sqrt{V}$, showing Ti d$_{z^2}$ character with a small O p$_z$ anitbonding contribution, for LDA and for an "enhanced" GW function which exaggerates the difference as explained in the text.

Fig. 3 (color online). SrTiO$_3$ LDA band structure for the O 2p upper valence bands and the low-lying conduction bands. The grey dashed lines are full LDA calculations, and the solid red lines are the Wannier interpolation. The dash-dotted OW and FW lines indicate the range of the outer and frozen energy windows used in the conduction-band MLWF construction.

Fig. 4 (color online). SrTiO$_3$ band structure comparing Wannier-interpolated LDA (solid red) and QSGW (dashed blue) upper valence and lower conduction bands. The open circles at the symmetry points denote the exact QSGW results on **k** mesh points.

Fig. 5 Second conduction band of solid Ar: squared modulus of the Bloch function along the direction (110) at $\mathbf{k} = (-1/8, -3/8, 1/4)$. White circles represent the location of the argon atoms.

Fig. 6 (color online). Full and interpolated solid Ar LDA conduction bands following the conventions of Fig. 3. The lower limits of the energy windows can be anywhere in the gap.

Fig. 7 (color online). Isosurface plots of QSGW MLWF's for solid Ar conduction bands at isosurface values $\pm 0.75/\sqrt{V}$. (a) One of six s-p-d hybrid-like functions pointing along the positive and negative Cartesian axes; (b) one of three t$_{2g}$ d-like functions.

Fig. 8 (color online). Comparison of Wannier-interpolated LDA and QSGW conduction bands for solid Ar, following the conventions of Fig. 4.

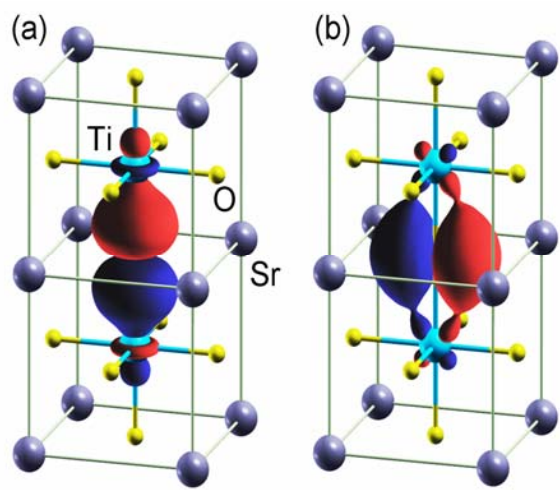

Fig. 1

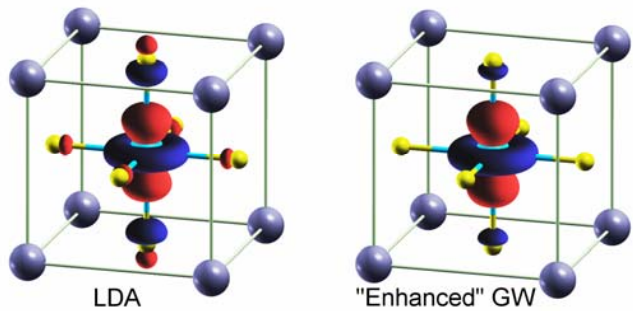

Fig. 2

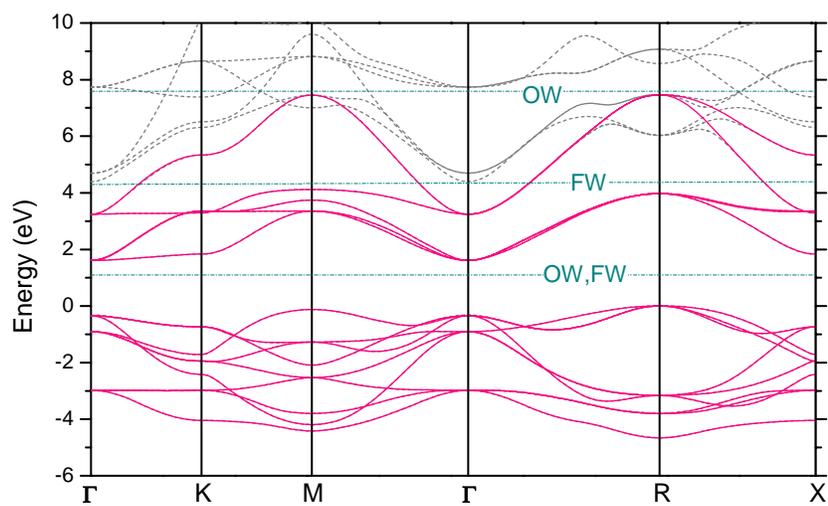

Fig. 3

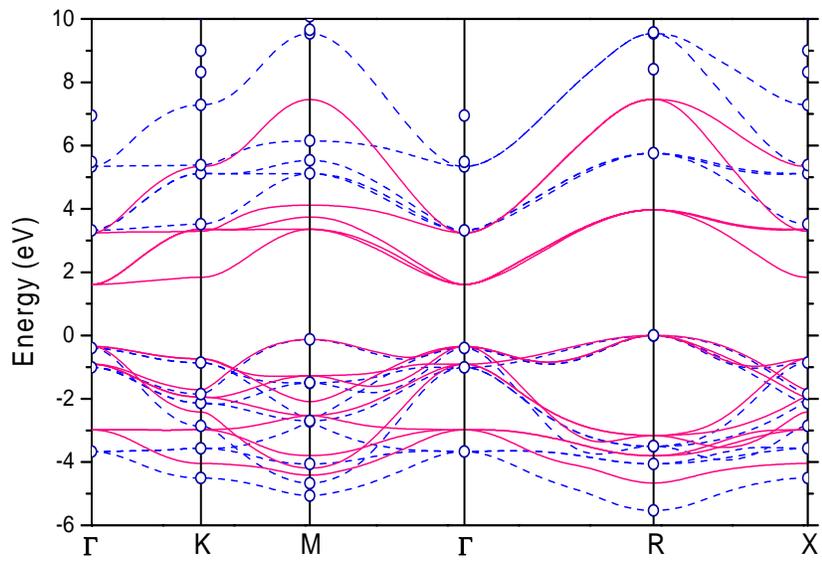

Fig. 4

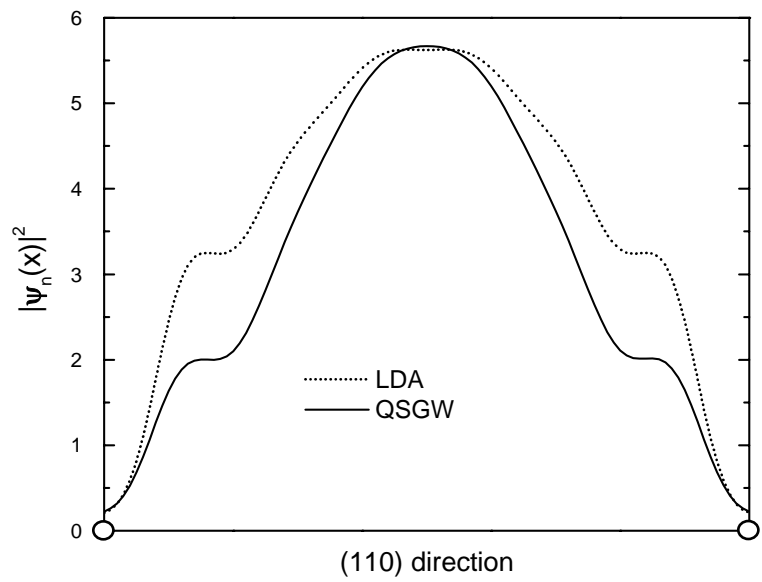

Fig. 5

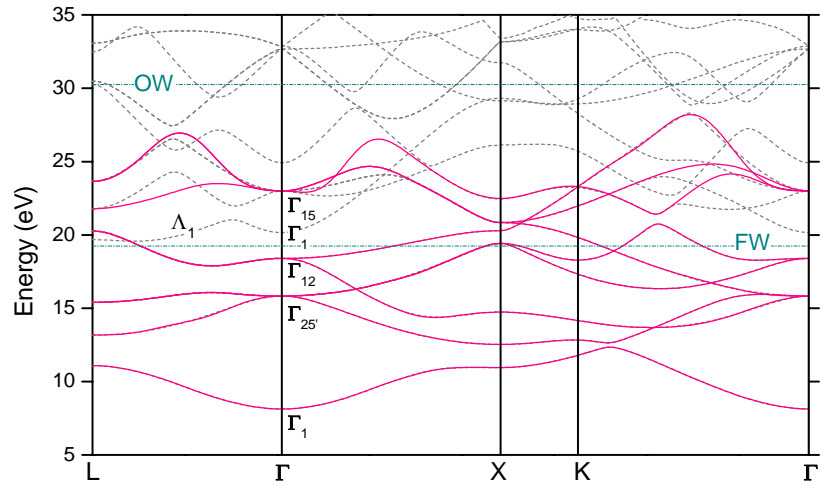

Fig. 6

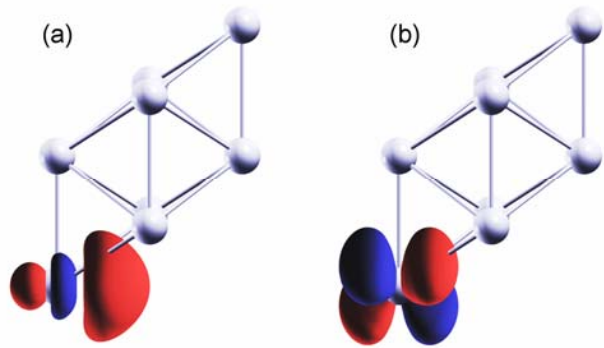

Fig. 7

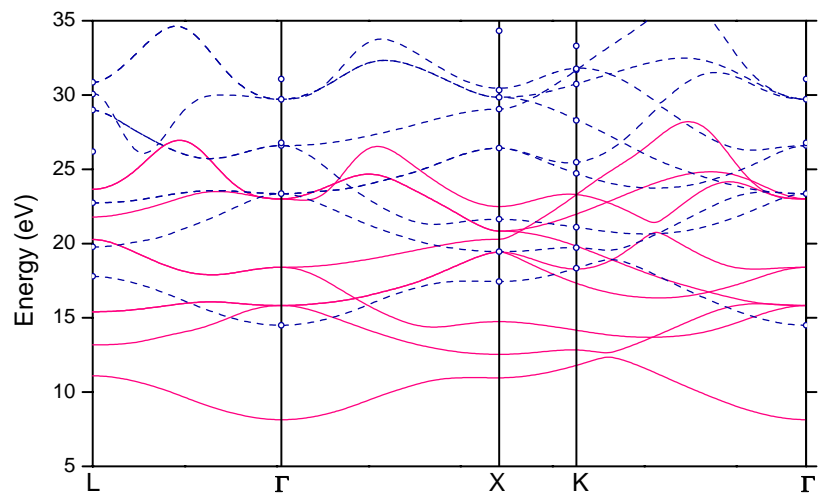

Fig. 8